\documentclass[aps,showpacs,preprint]{revtex4}
\usepackage{graphicx}
\usepackage{dcolumn}
\usepackage{bm}
\usepackage{epsfig}
\begin{document}
\title{Resistance fluctuations and $1/f$ noise in single crystalline Ni nanowires}

\author{Sudeshna Samanta\footnote[1]{email: sudeshna@bose.res.in}, M. Venkata Kamalakar\footnote[2] {email: venkat@bose.res.in} and A. K. Raychaudhuri\footnote[3]{email: arup@bose.res.in}}
\affiliation{Department of Materials Science, S.N.Bose National Centre for Basic Sciences, Block JD, Sector III, Salt Lake, Kolkata 700 098, West Bengal, India.}
\date{\today}
\begin{abstract}

\noindent
We measured the low frequency (10mHz $<f<$ 10Hz) resistance fluctuations (Noise) in single crystalline ferromagnetic Ni nanowires (diameter $\sim$ 35nm) in the temperature range 80K-300K. The noise spectral power shows $1/f$ dependence. The nanowires in an applied magnetic field show negative magetoresistance that saturates for $H \leq H_C$. The noise spectral power shows a reduction in low applied field and becomes field independent for $H \geq H_C$. This indicates that a part of the observed $1/f$ noise arises from magnetic origin. The magnetic part is associated with thermally activated domain wall fluctuations that couples to the resistance fluctuations.  
\end{abstract}


\maketitle

\noindent
Magnetic nanowires have potential applications in magneto-electric nanodevices~\cite{magnano}. With the development of advanced tools for nano-fabrications, the issue of fluctuations and noise in such nano -fabricated magnetic systems become important. In particular, in nanowires that have length much larger than the diameter, the nucleation of domain walls and its propagation between opposing magnetic domains (magnetization reversal) need to be understood as a basic problem because it is directly related to the stability and performance issues. In these wires the magnetization reversal process have been well studied in the nanosecond and microsecond range which is the time scale of the magnetization reversal either in a field and or by a current~\cite{magRevarsal}. However, in addition, there is a finite though small probability of thermally activated magnetization reversal and resulting domain nucleation and propagation. The kinetics of domain wall (DW) motion involve nucleation, pinning/de pinning and transit through the sample and thus has an element of stochastic process. It is thus expected that the domain wall kinetics may show low frequency fluctuations. This fluctuation can give rise to resistance fluctuation due to finite domain wall resistance and thus the study of low frequency noise can be an important tool to probe DW dynamics and low frequency magnetization fluctuation phenomena in nanowires (NW). In this letter we specifically investigate whether such a process occur in the nanowires. Low frequency resistance fluctuations like $1/f$ noise (that is in excess of Nyquist noise) can also arise from structural (non-magnetic) origin in a metallic system, in addition to that from the magnetic origin as stated above. We use an applied magnetic field to separate out the spectral power contributions from these two noise sources. The origin of the magnetic part of the resistance noise is traced to the fluctuations in DW and it has a direct correlation to the magnetoresistance (MR) that arises from the domain wall resistance. The experiment was carried out in very high quality single -crystalline nanowires of diameter 35nm in the temperature range 80K to 300K. We find that the noise from structural origin can be severely suppressed in such single crystalline and oriented samples (compared to polycrystalline samples) and we can identify the resistance noise from magnetic origin unambiguously. 

\indent
The ordered array of Ni nanowires (NiNWS) of average diameter $d \approx$ 35nm were synthesized by electrochemical deposition inside the pores of anodic alumina (AA) templates. The details of growth are given elsewhere~\cite{venkat1}. FIG~1(a) shows the Transmission Electron Microscope (TEM) image of the 35nm NiNWS. High resolution TEM (data not shown) shows absence of grain boundaries in these wires at least over large distances. FIG~1(b) also shows the X-ray diffraction (XRD) pattern of 35nm NiNWS with FCC lattice structure and shows alignment along (220) orientation. The NiNW's show a metallic resistivity ($\rho$) from 3K-300K. The residual resistivity of the wires was 0.065$\mu$Ohm.m (FIG~ 2(inset)). A detailed analysis of $\rho$ NiNWs down to 3K (the regime of residual resistivity) showed that the electron mean free-path $\approx1.1d$~\cite{venkat1}. This implies absence of significant grain boundary or internal defect scattering within the bulk of the wires. The residual resistivity thus predominantly arise from nearly diffused scattering of electrons from the surface. We see below that a very small fraction ($<1\%$) of the resistivity arise from magnetic DWs which in turn are the likely cause for the low frequency $1/f$ noise of magnetic origin in these materials. The nanowires ($d=$ 35nm) show nearly rectangular $M-H$ curve with field applied along the wire axis taken at $T$ = 300K (FIG~1(c)). The data are taken in an array by retaining the wires in a template. The $M-H$ curve shows that the saturation magnetization is reached at H$<$0.25 Tesla and the coercivity $H_{C}\approx$ 0.1 Tesla. The retention ratio $M_{R}/M_{S}\approx$ 0.90. $H_{C}$ has a very shallow dependence on temperature and increases on cooling. This has been observed before and explained using a localized nucleation model of thermally activated magnetization reversal that decreases the $H_C$ from the "ideal" value and also makes $H_C$ reduce with increasing $T$~\cite{paulus,shellmyer}. 

\indent
The resistance and noise measurements were carried out in the temperature range 80K-300K by retaining the NW inside the templates, where each nanowire was separated by the insulating alumina intervening medium. The electrical contacts were prepared by low-noisy silver epoxy on about 1mm$\times$1mm contact area. The same contacts were used for resistivity and noise. The details of data acquisition and digital signal processing (DSP) are described elsewhere~\cite{ghosh}. The resistance fluctuation shows up as a voltage noise in the current biased sample. This technique of noise measurement has been used in the past for investigation in NW of silver (Ag) and copper (Cu), and it ensures that contact noise does not contribute~\cite{bid}. The digitized voltage signal was processed by using DSP technique and the experiment was carried out over 3 decades of frequency from 10mHz-10Hz. The spectral power of the back ground noise including the Nyquist noise was $\approx$ 2$\times$10$^{-20}V^{2}/$Hz. FIG~2 illustrates the power spectral density, plotted as $S_{V}(f)/V^{2}$, as a function of $f$ at two representative temperatures. $S_{V}(f)/V^{2}$ increases on heating. However, it has a weak temperature dependence below 120K. The measured $S_{V}(f)$ is substantially higher than any background noise which has been subtracted out. The spectral power over the whole temperature range shows that $S_{V}(f)\propto 1/f^{\alpha}$ where $\alpha$ is close to $\approx$ 1-1.2, confirming the essential $1/f$ character of the observed noise. The probability distribution function of the observed resistance fluctuation is Gaussian which ensures that the fluctuators producing the noise are statistically independent of each other. The relative mean squared resistance fluctuation $<(\Delta R)^{2}>/R^{2}$ can be obtained by integrating the $S_{V}(f)/V^{2}$ over the band width of measurement. The value of $<(\Delta R)^{2}>/R^{2}$ is independent of the number of nanowires in the array. Each nanowire thus acts as an independent noise source. We have compared the $<(\Delta R)^{2}>/R^{2}$ of the nanowire with that obtained by a similar experiment on a polycrystalline Ni wire of diameter $\sim$ 50$\mu$m (referred as "bulk"). We find that over the whole temperature range the $<(\Delta R)^{2}>/R^{2}$ for the nanowire array is at least a factor of 3 less than that seen in the "bulk". The larger noise in the "bulk" wire has been shown also in FIG~2 where we show representative spectral power. We find in polycrystalline NW of similar dimension the spectral noise in the same range can be more than 2 orders more. The observed resistance fluctuation thus has a component from structural (non-magnetic) origin and it is much less in NW of the higher crystalline quality. The observed $S_{V}(f)$ also contains a magnetic part. In the FIG~2 we show the spectral power measured under a magnetic field $H$= 0.12 Tesla $\simeq H_{C}$ at 80K. The reduction of $S_{V}(f)/V^{2}$ in an applied field is very clear. From the data on spectral power in zero field and in an applied field we could separate out the magnetic contribution from that arising from the structural origin. Discussion on the resistance fluctuation arising from structural origin is beyond the limited scope of the paper. In the rest of the paper we discuss the exact magnitude of the magnetic part of the noise, its correlation to the magnetoresistance and its likely physical origin.

\indent
In a ferromagnetic NW with diameter $d <\sqrt2 d_W$ and length $>> d_W$ ($d_W$ = domain wall width), the wire has one dimensional magnetic behavior with all the moments aligned parallel or antiparallel to one direction. For Ni, $d_{W}\sim$ 35-40nm~\cite{paulus,shellmyer}. The 35nm wires can thus be regarded as being one-dimensional. Also, the shape anisotropy energy being at least one order more than the magneto-crystalline anisotropy in Ni, the easy axis is along the axis of the wire~\cite{paulus,shellmyer}. Extensive study of temperature dependent hysteresis loop in NiNW of this diameter range, temperature dependence of $H_C$ and slow dynamics seen in magnetic viscosity, have established that the magnetization reversal in these wires takes place not in unison or by curling modes but by nucleation of domains of opposite spins that are localized in small effective volume~\cite{paulus}. The dynamics of the localized nucleation phenomena is controlled by the dynamics of the domain walls so created. In zero field the probability of reversal by thermal energy alone is low, given the height of the activation barriers, however, it is still finite and give rise to slow spin dynamics. 

\indent
We show below that the low field MR which due to finite resistance of the DWs can couple to low frequency DW dynamics and can show noise that is correlated to the MR. In FIG~3(a) we show the magnetoresistance of the nanowire taken at 80K. The reduction of the resistance in magnetic field is small though distinct and it saturates for $H> H_{C}$. Beyond 0.04 Tesla, the field dependence is shallow. We connect this resistance change at low field predominantly arising from DWs which in such ferromagnetic wires can have a positive resistance of around $\sim$ 1-2$\%$~\cite{percentResist}. When the wire is fully magnetized there is no DW in the wire (uniformly magnetized) and the resistance mainly arises from non-magnetic sources. In zero fields (which we start as a demagnetized virgin state) there are domains present in the system. Thus the negative MR seen in the samples occur predominantly due to positive resistance of the DWs. In FIG~3(b) we also show the measured $<(\Delta R)^{2}>/R^{2}$ as a function of $H$ at the same temperature. It can be seen that when the resistance change nearly saturates, the fluctuation also reaches a field independent value, which we assign to arise from structural origin alone. The main change in resistance as well as the noise occurs for $H \leq 0.05$ Tesla. The above correlation of the MR and the magnetic field dependence of the resistance fluctuations are seen over the whole range of temperature. Thus, the likely magnetic source of resistance fluctuation in a narrow ferromagnetic nanowires ($d <\sqrt 2 d_W$) can be linked to the DW resistance and their dynamics that accompanied the thermally activated magnetization reversal. The separation of the magnetic and non-magnetic part of the power spectrum on application of field allows us to write, $S_{V}(f)/V^{2}$ = $[S_{V}(f)/V^{2}]_{m}$ + $[S_{v}(f)/V^{2}]_{nm}$, where subscripts $m$ and $nm$ refer to magnetic and non-magnetic origins respectively. The $[S_{V}(f)/V^{2}]_{nm}$ is defined as the $H$ independent value of $S_{V}(f)/V^{2}$ for $H\geq H_C$. The non-magnetic and magnetic parts of $S_{V}(f)/V^{2}$ are marked by a dashed line and by an arrow in FIG~3(b) respectively. In FIG~4 we have plotted $[S_{V}(f)/V^{2}]_{m}$ so obtained for 4 different frequencies as function of temperature to bring out the activated nature of the fluctuators that give rise to the noise.

\indent
The observed magnitude of $[<(\Delta R)^{2}>/R^{2}]_{m}$ is $\sim 10^{-12}$. If all the DWs that contribute to the resistance would have contributed to the noise in the time scale of the measurements, the fluctuation $[<(\Delta R)^{2}>/R^{2}]_{m}$ would be $\approx 10^{-4}$. This is much larger than what we observe. This is because what we observe within our detection band width ($f_{min}$ = 10$^{-2}$Hz to $f_{max}$ = 10Hz) are those fluctuations that occur within the time scale $1/2\pi f_{max}$ sec $<t<$ $1/2\pi f_{min}$ sec. Fluctuations outside this time window will not be recorded by us. Thus the experiment captures a small fraction of the total fluctuators. This is expected as stated above that the magnetization reversal and associated domain dynamics due to the stochastic nature can occur over a substantial time scale~\cite{magRevarsal}. The $1/f$ nature of the power spectrum occurs from this long distribution of time scale. In the arrays that we are making measurements there are nearly 100 nanowires in parallel, each of which acts as an independent source. This leads to a distribution of the relaxation time and the $1/f$ nature of the power spectrum.

In conclusion, we report the first measurement of resistance fluctuation in ferromagnetic NW with diameter that would make the spin-system one-dimensional. The noise spectral power shows clear $1/{f^\alpha}$ behaviour. The power spectrum contain a clear contribution that has a magnetic origin. We find that the low frequency fluctuations can arise DW fluctuations.

The authors thank the Department of Science and Technology (DST), Govt. of India for a sponsored project as Unit. M.V.K thanks CSIR, Govt. of India for financial support. Partial financial support from CSIR as a sponsored research project is gratefully acknowledged.

\newpage
\section{\large\bf Figure Captions}


\noindent {\bf FIG~1:} (Color online) (a) The TEM image of 35nm Ni NW and its (b) XRD pattern shows a (220) peak. (c) $M-H$ curve at 300K taken on NW array with magnetic field parallel and perpendicular to the wire axis.\\
\noindent {\bf FIG~2:} (Color online) The observed power spectra (background subtracted) in NiNWS along with a 50$\mu$m “bulk” wire at two representative temperatures (80K and 300K). The $1/f$ nature of power spectra is noted. The graph also shows the field dependence of the power spectrum on NiNWS at 80K. The upper inset shows the resistivity measured as a function of temperature. \\
\noindent {\bf FIG~3:} (Color online) Magnetic field dependence of (a) resistance of NiNWS and (b) $<(\Delta R)^{2}>/R^{2}$ at 80K.\\
\noindent {\bf FIG~4:} (Color online) Temperature dependence of the magnetic part of $S_{V}(f)/V^2$ (see text) at different temperatures. The activated nature can be distinguished from the broad peak as a function of tempereature.\\
\newpage
\begin{figure}
{\resizebox{6.0in}{!}{\includegraphics{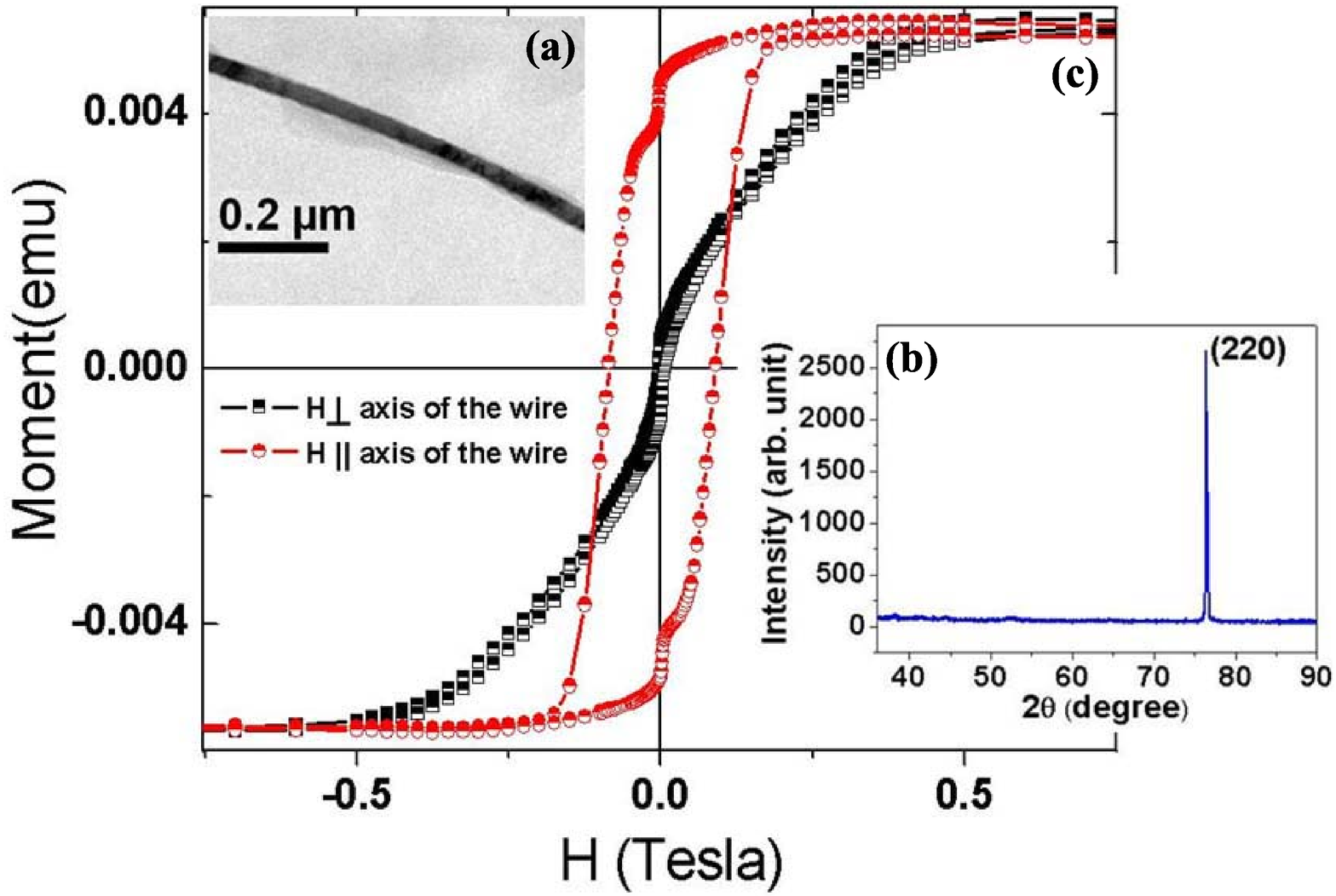}}}

\caption{}
\end{figure}

\begin{figure}
{\resizebox{6.0in}{!}{\includegraphics{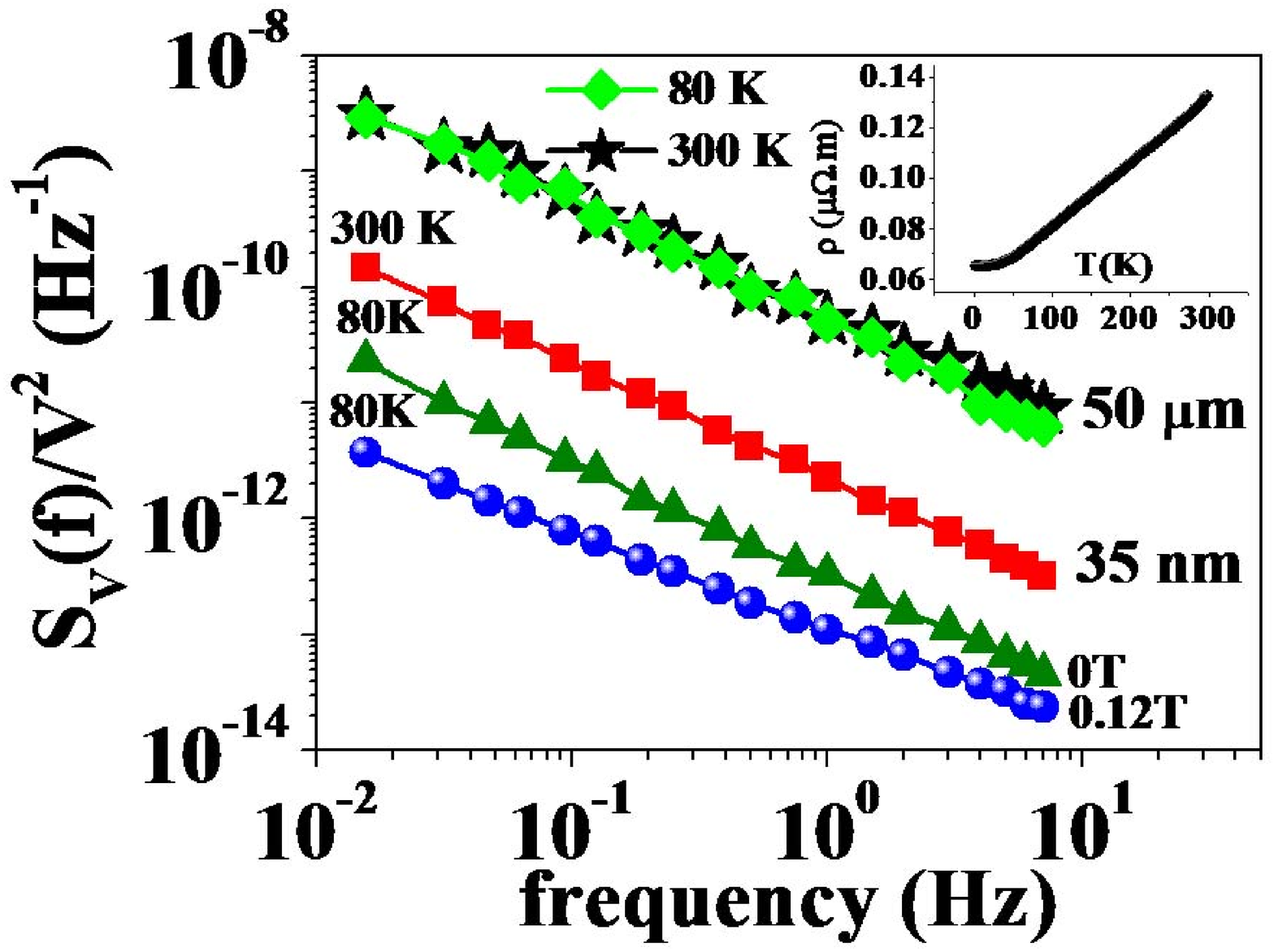}}}

\caption{}
\end{figure}

\begin{figure}
{\resizebox{6.0in}{!}{\includegraphics{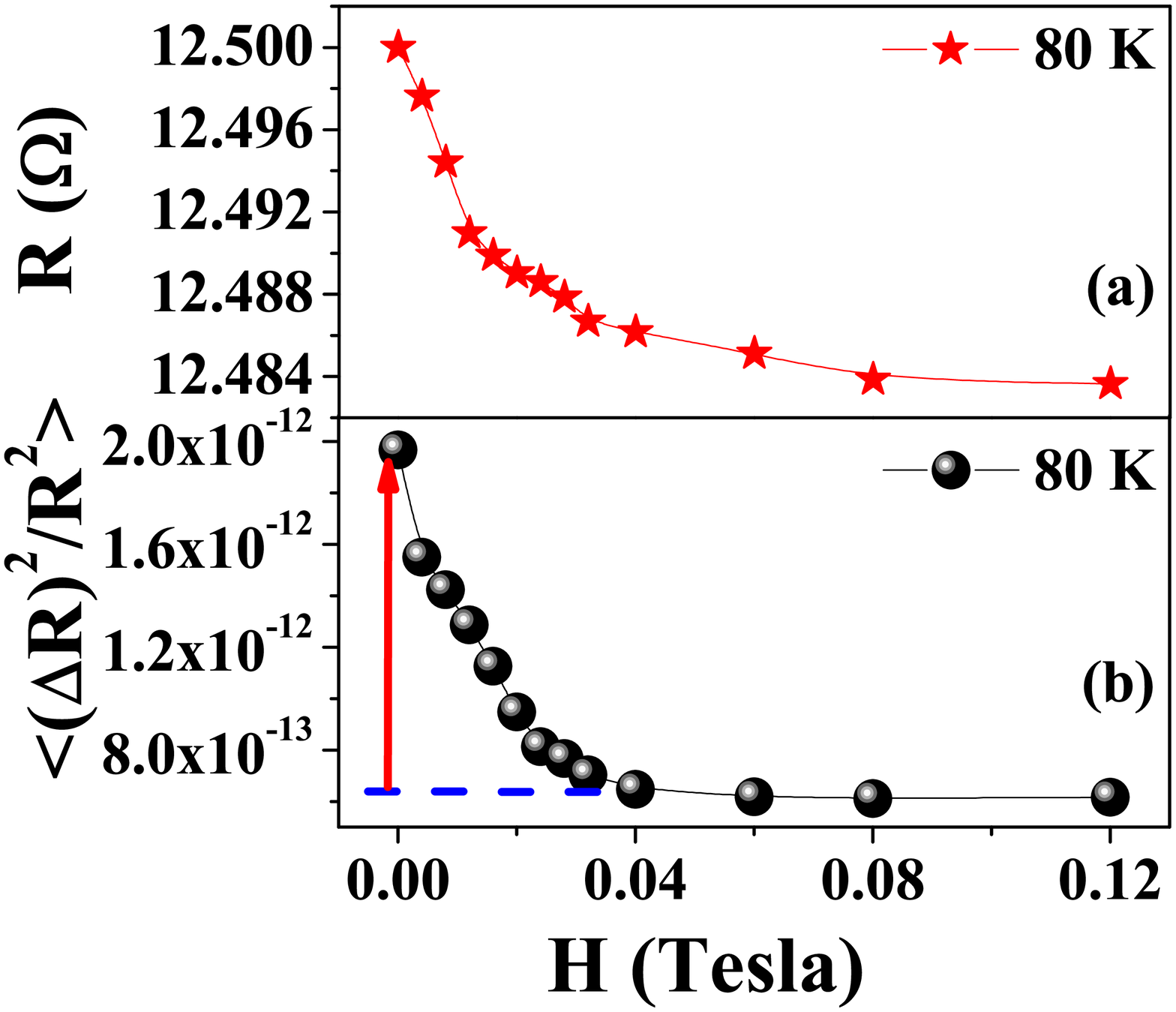}}}

\caption{}
\end{figure}

\begin{figure}
{\resizebox{6.0in}{!}{\includegraphics{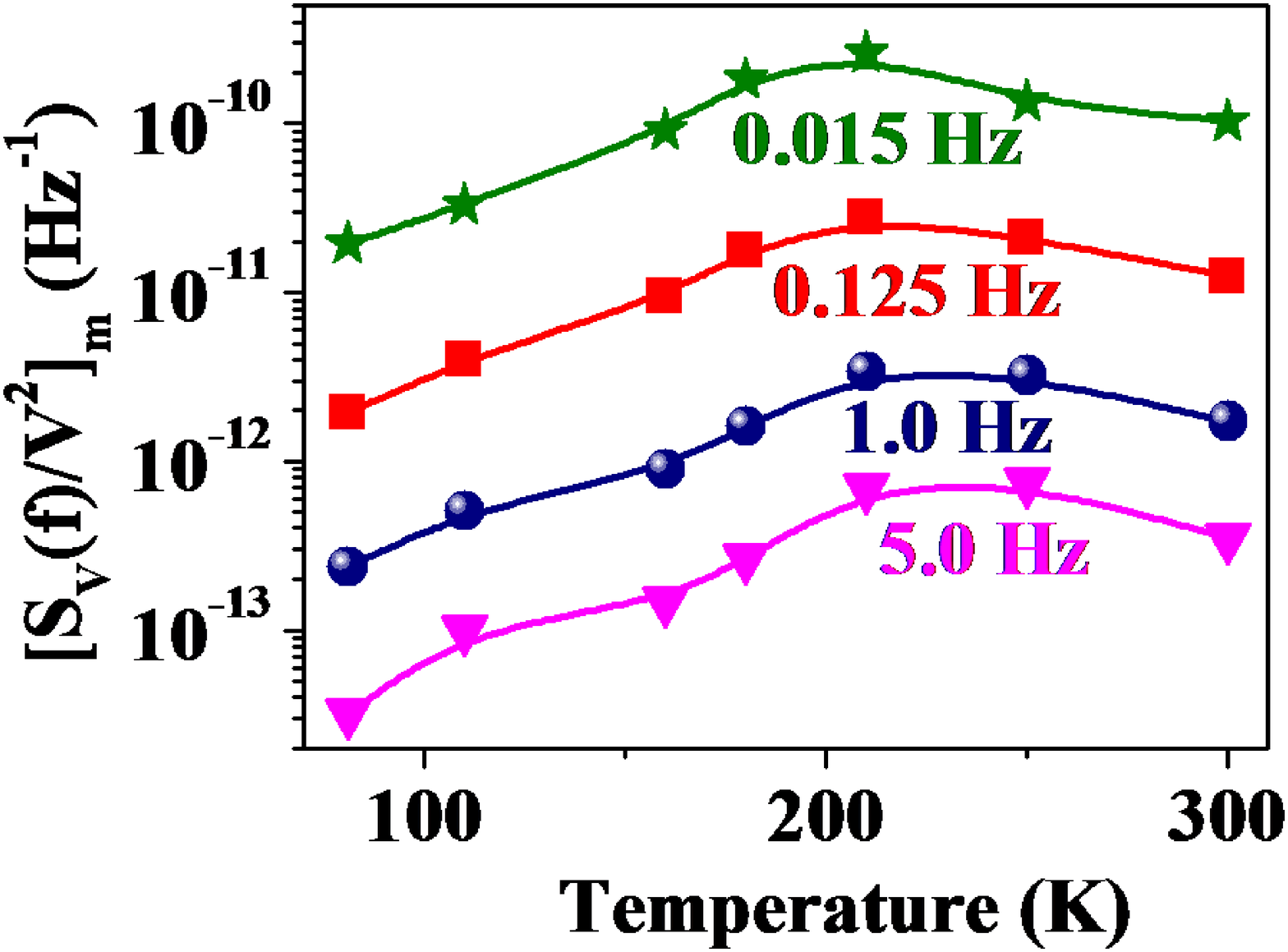}}}

\caption{}
\end{figure}
\end{document}